# Disordered vs. Short-Range-Ordered Complexions: Consequences for Grain-Boundary-Mediated Plasticity in Nanocrystalline Al-Ni Alloys

Frederic Sansoz[a,b,*] and Eve-Audrey Picard[a]

[a]Department of Mechanical Engineering, The University of Vermont, Burlington, Vermont 05405, USA

[b]Materials Science Program, The University of Vermont, Burlington, Vermont 05405, USA

*To whom correspondence should be addressed. Tel: (802)656-3837. Email: frederic.sansoz@uvm.edu

## ABSTRACT

Disordered grain-boundary (GB) complexions in thermally stable nanocrystalline alloys are commonly assumed to be structurally uniform, yet their amorphous nature makes them susceptible to local short-range ordering (SRO). The influence of local SRO on GB-mediated plasticity mechanisms in such complexions remains poorly understood. This article employs large-scale Monte Carlo and molecular dynamics simulation to address this gap through simulations of nanocrystalline Al-Ni alloys at two Ni concentrations, 2 at.% and 4 at.%. Annealing at 913 K produces thick uniform disordered intergranular film complexions, while annealing at 378 K produces semi-disordered complexions containing FCC-type and BCC-type SRO. These two complexion states produce fundamentally different mechanical responses. Amorphous complexions act as dislocation sinks, promoting homogeneous plasticity through shear transformation zones, but at the cost of intense shear localization and lower strength. SRO complexions generate reduced shear localization and higher strength but also promote heterogeneous stress concentrations across the GB network, regardless of Ni concentration. This contrast reflects a fundamental shift in governing mechanism, from shear-transformation-zone-controlled behavior in disordered complexion alloys to GB-stress-heterogeneity-controlled behavior in SRO complexion alloys. These findings highlight the potential of complexion engineering to tailor the mechanical properties of nanocrystalline materials.

## I. INTRODUCTION

Grain-boundary (GB) complexions are thermodynamically stable phases that exist exclusively at crystalline interfaces [1] and are responsible for enhancements in thermal stability [2], radiation damage tolerance [3,4] and mechanical properties [5] in nanocrystalline alloys. Complexions naturally transition from ordered to disordered phases depending on the GB chemistry, solute concentration, and thermodynamic conditions [6-8]. In nanocrystalline Al, a 2 at.% Ni addition produces ordered interfaces when the powder-consolidated Al-Ni alloy is slowly cooled, whereas thick amorphous GB complexions form upon firing followed by rapid quenching [9]. In contrast, a 2 at.% Y addition consistently produces amorphous GB complexions of varying thicknesses [9], underscoring how sensitively complexion structure depends on both alloy chemistry and thermal history.

The nature of these complexions has profound consequences for mechanical behavior. Amorphous intergranular complexions exhibit more uniform solute distribution across the GB network than heterogeneous segregation, as demonstrated in sputtered Ni-Zr alloys by Schuler and Rupert [6]. This uniformity gives rise to thick disordered GB regions that critically influence deformation mechanisms [10]. Experimental and atomistic simulation studies [6,8,9,11] consistently show that thick amorphous GBs better absorb dislocations and slow their propagation, thereby enhancing ductility in nanocrystalline alloys [12,13]. This effect is attributed to the capacity of the GB network to store excess free volume as a function of thickness, which controls localized plasticity and stress accommodation at the nanoscale [14]. Further evidence stems from the glass-forming nanocrystalline $Al_{85}Ni_{10}Ce_5$ alloy studied by Balbus et al. [15], where strong shear localization was observed following plasticity onset under indentation. They found that Ni and Ce solute segregation at GBs after low-temperature annealing significantly reduced shear localization and increased strength, due to phase formation along the interfaces.

The internal structure of disordered GB complexions could be assumed to be homogeneous. However, early literature on metallic glasses strongly suggests the formation of local short-range ordering (SRO) inside amorphous phases at equilibrium [16]. Similarly, at the atomic scale, the chemically and structurally random structure of GB complexions makes them susceptible to local SRO [17-19], whose influence on GB-mediated plasticity in nanocrystalline alloys remains poorly understood. The present article addresses this gap by employing Monte Carlo and molecular dynamics (MC/MD) simulations to investigate how local SRO in amorphous intergranular complexions impacts on the plasticity of nanocrystalline Al-Ni alloys at two Ni concentrations, 2 at.% and 4 at.%.

## II. COMPUTATIONAL METHODOLOGY

MC/MD simulations were performed with the Large-scale Atomic/Molecular Massively Parallel Simulator (LAMMPS) [20] using the embedded-atom-method potential for Al-Ni pair interactions by Mishin et al., which was fitted on *ab initio* calculations [21]. The simulations were performed in a variance-constrained semi-grand-canonical (VC-SGC) ensemble [22] following the same methodology as previously described in reference [23]. MC trials were run with constant parameters T, $c_0$, $\Delta\mu_0$, and $\kappa$ which are, respectively, the annealing temperature, the total solute concentration, the initial chemical potential difference between Al and Ni elements, and the variance constraint parameter. The annealing temperatures were set to 913 K and 378 K, corresponding to the maximum solubility temperature for Al-Ni [24], and a lower reference temperature, with homologous temperatures of $0.978 \cdot T_m$ and $0.405 \cdot T_m$, respectively, based on the melting point of pure Al, $T_m = 934$ K [24]. For this study, we used $\Delta\mu_0 = 1.790$ eV at T = 378 K and 1.990 eV at 913 K, and $\kappa = 2000$. The concentration $c_0$ was fixed at 2 at. % and 4 at.% Ni. In the following, these alloys will be referred to as Al-2Ni and Al-4Ni, respectively.

Each MC cycle consisted of a total number of trial moves equal to the number of atoms in the system and was separated by 1,000 MD steps at the same temperature, which was iterated 600 times. The MD was performed in the isothermal-isobaric (NPT) ensemble using a Nose-Hoover temperature thermostat and a zero pressure barostat. The MD timestep was 0.001 ps. The chemical potential difference was rescaled after every 10,000 MD steps to ensure convergence to the target solute concentration $c_0$ over the total duration of simulation, 600 ps.

MC/MD was applied to a face-centered-cubic (FCC) Al polycrystal of approximately 13 million atoms created with the software Atomsk [25]. The simulation box measured 60 nm in the x, y, and z directions with periodic boundary conditions and contained 27 grains of random orientations and positions, corresponding to an average grain size of ~24 nm. After MC/MD, the polycrystal was cooled to 300 K at 2 K $ps^{-1}$ under zero pressure using NPT integration with a timestep of 0.002 ps, followed by equilibration at 300 K for 100 ps. Tensile deformation was then applied along the x-axis at a constant engineering strain rate of $10^8$ $s^{-1}$ up to 10% strain, while maintaining zero pressure in the y and z directions using NPT integration. A timestep of 0.005 ps was used during deformation. Total strain was computed from the applied strain rate and elapsed simulation time, and tensile stress was obtained from the box pressure along the x-axis. Simulation snapshots were saved every 5 ps. Local structural ordering and chemical ordering by polyhedral template matching analysis, using an RMSD cutoff of 1.13, grain segmentation analysis, and atomic von Mises shear strain analysis, using a cutoff radius of 3.3 Å, were performed in the atomistic visualization software OVITO [26]. Atomic-scale von Mises stress distributions were computed from the LAMMPS dumps by averaging each component of the virial stress tensor over 2.5 ps in time, corresponding to 500 MD steps, and in volume within a sphere of 6 Å radius.

The radial distribution function (RDF) for Ni-Ni pairs was used to characterize the short-range order of Ni solute atoms segregated to the GB network. The RDF provides insight into dominant interatomic distances in Ni-Ni pairs through the positions of its peaks [27]. The RDF peaks were compared to those of a perfect cubic Ni lattice where the first peaks correspond to the nearest neighbor (NN) distances. For an FCC lattice of constant $a$, the first four NN distances are $a\sqrt{2}/2$, $a\sqrt{3}/2$, $a$, and $a\sqrt{2}$, respectively. Using $a$ = 3.52 Å for FCC Ni [28], the corresponding first NN distances are 2.42 Å, 2.80 Å, 3.52 Å, and 4.98 Å for FCC Ni, respectively.

The strain localization factor (LF) by Duan et al. [29] was calculated on each polycrystal, by using the formula:

$$LF = \frac{1}{V} \sum_{i<j,\varepsilon_i>\bar{\varepsilon},\varepsilon_j>\bar{\varepsilon}} \frac{(\varepsilon_i/\bar{\varepsilon} - 1)(\varepsilon_j/\bar{\varepsilon} - 1)}{r_{ij}}, \quad (1)$$

where $\boldsymbol{\varepsilon_i}$ and $\boldsymbol{\varepsilon_j}$ are the von-Mises shear strain at atom positions indexed with $i$ and $j$, respectively, $\bar{\boldsymbol{\varepsilon}}$ the average shear strain, $\boldsymbol{r_{ij}}$ the distance between positions $i$ and $j$, and $V$ the sample volume. This factor is used to measure the intensity of strain localization in the whole polycrystal. A cutoff of 12 Å was used for computational efficiency. As explained by Duan et al., we also accounted for the difference in average strain by comparing the modified localization factor (MLF) calculated as the product of LF and average shear strain. LF and MLF calculations were coded as a python module and added to the OVITO pipeline.

## III. RESULTS

### III.1 Complexion structures

Figure 1 shows a pre-deformation snapshot of a nanocrystalline Al-4Ni polycrystal following MC/MD simulation at 378 K. Most Ni atoms segregate homogeneously to the GBs, though segregation is incomplete, as some Ni solute clusters persist within the grain interiors. The nearly identical grain size at different temperatures results from the short MD simulation time (600 ps) performed during MC/MD annealing step and strong GB pinning effects when Ni solute atoms segregate to GBs. The simulated Al-Ni phase diagram in Figure 2 reveals that annealing at 913 K eliminates a majority of intragranular Ni clusters, consistent with this temperature near the maximum solubility limit of 934 K. At 913K, the GB thickness increases noticeably from 2 at. % Ni to 4 at. % Ni. The average GB thickness respectively doubles from 8.2 Å ± 4.2 Å to 16.5 Å ± 8.1 Å, while the GB atom fraction rises from 12.4% to 23.9% (see Table 1).

Table 1: Local structural and chemical ordering of all Ni atoms in each pre-deformed model after MC/MD thermal annealing followed by cooling to 300 K from polyhedral template matching analysis.

| | | | **Local structural Ni ordering (%)** | | | | **Local chemical Ni ordering (%)** | | | | |
|---|---|---|---|---|---|---|---|---|---|---|---|
| **Alloy** | **MC/MD temperature** | **GB atom fraction (%)** | **No ordering** | **FCC** | **BCC** | **ICO** | **No ordering** | **Pure Ni** | **$L1_0$** | **$L1_2$** | **B2** |
| Al-4Ni | 913 K | 23.9 | 98.6 | 1.4 | 0.0 | 0.0 | 97.4 | 0 | 0 | 2.6 | 0 |
| | 378 K | 15.1 | 84.8 | 9.5 | 5.6 | 0.0 | 99.2 | 0 | 0.2 | 0.6 | 0 |
| Al-2Ni | 913 K | 12.4 | 97.6 | 2.3 | 0.1 | 0.0 | 96.5 | 0 | 0 | 3.5 | 0 |
| | 378 K | 9.1 | 95.4 | 3.1 | 1.5 | 0.0 | 99.1 | 0 | 0.1 | 0.8 | 0 |

At 2% Ni and 913 K, the resulting structure closely resembles the experimental transmission electron microscopy (TEM) images of GB complexions reported by Lei et al. [9] in nanocrystalline fired-and-quenched Al-2Ni, although their quenched alloy also exhibited some $Ni_3Al$ nanoprecipitation that was not observed in our simulations. At 4% Ni and 913 K, the thicker GB region is more consistent with amorphous interfacial films observed by Wu et al. [30] via TEM in nanocrystalline Al-4Ni-3Y alloy, where pure Al nanograins were surrounded by nano-sized metallic glass shells. Glass-like GB complexions are known to form in alloys with good glass-forming ability, wherein solute atoms are rejected to the GBs due to low matrix solubility [6].

The amorphous character of the GB complexion in Al-4Ni at 913 K is confirmed by the Ni-Ni RDF curves in Figure 3. At 913 K, the RDF curve exhibits broad, diffuse peaks between the first and fourth NN distances, which is characteristic of metallic glasses [31]. The occurrence of a fourth peak is not surprising because it corresponds to a Ni-Ni separation equal to twice the minimum Ni-Ni distance at the first peak. The absence of additional peaks between these features indicates a homogeneous and random Ni arrangement at GBs.

In contrast, at 378 K, the Al-4Ni RDF displays three additional peaks between the first and fourth NN distances, indicative of local SRO. One peak falls near the second FCC Ni NN distance; however, the remaining two peaks, located at approximately 3 Å and 4.1 Å, do not correspond to any FCC Ni NN distances, suggesting a non-FCC ordering. These positions are instead consistent with BCC-type SRO at the GBs, as shown hereafter.

Figure 4 compares the atomic-scale GB complexion structure in Al-4Ni at 913 K and 378 K using the polyhedral template matching algorithm, which can distinguish between different structural and chemical ordering types [32]. At 378 K, the interface contains structurally ordered Ni atoms of both FCC and BCC character, whereas the same interface at 913 K shows no detectable SRO, which agrees with the RDF in Figure 3. Based on Figure 4(b) and our own observations, the SRO corresponds to nanoscale clusters of less than 10 atoms.

To identify the SRO type, Table 1 presents the local structural and chemical ordering of all Ni atoms in each pre-deformed model after MC/MD thermal annealing followed by cooling to 300 K. The results show that as the MC/MD temperature is decreased to 378 K, the GB atom fraction decreases and local structural Ni ordering of both FCC and BCC type increases. It should also be noticed that no icosahedral order was found, suggesting that the GB region differs from the icosahedral SRO of standard amorphous alloys [33]. As the MC/MD temperature is increased, the GB region becomes more than 98% disordered, while the remaining Ni order is due to interface Ni atoms adjacent to FCC grains. In addition, Table 1 shows almost no local chemical Ni ordering in all models.

Taken together, these results indicate that at 4% Ni, GB segregation produces a thick structurally disordered intergranular complexion at 913 K, whereas at 378 K it produces a semi-disordered intergranular complexion containing BCC-type and FCC-type SRO. At 2% Ni, the complexion type follows the same temperature dependence; however, GB thickness is thinner due to smaller GB atom fraction. As shown in Figure 2 and Table 1, the primary distinction between temperatures and Ni concentrations lies in the degree of intragranular Ni clustering and GB SRO, both of which decrease with either increasing temperature or decreasing Ni concentration. Overall, Table 1 reveals that the local SRO of the GB region is structural rather than chemical.

### III.2 Plastic deformation behavior

Figure 5 shows the tensile behavior of nanocrystalline Al-2Ni and Al-4Ni with amorphous GB complexions (annealed at 913 K). Both alloys exhibit significant strengthening relative to pure nanocrystalline Al annealed at 913 K, with Al-2Ni achieving higher strength than Al-4Ni up to Position 1 in Figure 5. At 3% strain (Position 1), deformation is dominated by GB-induced plasticity, which is more pronounced in Al-4Ni than in Al-2Ni, with no lattice dislocation activity observed in either alloy. This is consistent with the larger fraction of disordered GB complexion in Al-4Ni (913 K, Table 1). At higher strains corresponding to Position 2 (6%) and Position 3 (9%), the two alloys behave nearly identically, as deformation transitions from GB-mediated plasticity to crystal plasticity driven by dislocation emission from GBs.

In contrast, Figure 6 shows that nanocrystalline Al-2Ni and Al-4Ni with SRO GB complexions (annealed at 378 K) exhibit nearly identical stress-strain curves throughout. Both alloys reach the highest maximum stress of all simulations at Position 1 (3% strain), followed by an abrupt stress drop at Position 2 (6% strain) and a stable plastic flow regime at Position 3 (9% strain). This response is consistent with the strain-localized shear band behavior described by Steif et al. [34]. Notably, the deformation snapshots in Figures 5 and 6 show that shear strain is already localized to the GB regions at this early stage (Position 1), suggesting that SRO configurations at GBs influence strain localization from the onset of deformation.

In Table 2, we quantified the severity of strain localization between each model at different stages of deformation using LF and MLF analysis. The results highlight two conclusions. First, the difference in strain localization between complexion types is more pronounced at the early deformation stage before dislocation emission (Position 1, Figures 5 and 6) than during dislocation-mediated plasticity at larger strain (Position 2, Figures 5 and 6), even though shear localization overall increases with average strain. Second, at Position 1, shear localization is more severe in the disordered complexions (annealed at 913 K) than in the SRO complexions (annealed at 378 K).

Table 2: Strain localization factor (LF) and modified localization factor (MLF) at strain positions 1 and 2.

| | | | **Total strain 3% (Position 1)** | | | **Total strain 6% (Position 2)** | | |
|---|---|---|---|---|---|---|---|---|
| **Alloy** | **MC/MD temperature** | **GB complexion Type** | **Average shear strain** | **LF ($Å^{-4}$)** | **MLF ($Å^{-4}$)** | **Average shear strain** | **LF ($Å^{-4}$)** | **MLF ($Å^{-4}$)** |
| Al-4Ni | 913 K | Disordered | 0.060 | 0.775 | 0.046 | 0.115 | 1.321 | 0.151 |
| | 378 K | Short-range ordered | 0.039 | 0.161 | 0.006 | 0.097 | 1.622 | 0.158 |
| Al-2Ni | 913 K | Disordered | 0.047 | 0.653 | 0.031 | 0.104 | 1.206 | 0.125 |
| | 378 K | Short-range ordered | 0.038 | 0.173 | 0.007 | 0.096 | 1.199 | 0.116 |

### III.3 Atomic stress distributions

Since the most significant differences in deformation mechanism occur before the maximum stress, we focus on the atomic von Mises stress distribution at Position 1 (3% strain) within the GB interface shown in Figure 4. This GB region is of particular interest because this site shows different shear strain distributions between each complexion alloy (Figures 5 and 6).

As shown in Figure 7(a), stresses in the GB regions of Al-4Ni with disordered GB complexion are nearly fully relaxed, indicating effective stress accommodation by the thick amorphous interfacial film. In Al-2Ni with amorphous complexion in Figure 7(b), however, localized stress concentrations appear within the GB network (highlighted by arrows), which may account for the strength difference between the two amorphous complexion alloys observed in Figure 5.

For both alloys with SRO complexions, Figure 8 shows that the GB network exhibits repeated microscopic stress concentrations (highlighted by arrows) that are absent in Al-4Ni at 913 K with its thick amorphous GB complexion (Figure 7). Overall, these findings suggest that SRO complexions are responsible for a fundamental shift in the local GB stress distribution and associated GB plasticity mechanisms.

## IV. DISCUSSION

This discussion focuses on the relationship between local SRO in GB complexions and GB-mediated plasticity mechanisms in nanocrystalline alloys. At 913 K, near the solubility limit, the present study shows that nanocrystalline Al-Ni alloys form thick disordered intergranular complexions whose thickness is strongly influenced by Ni concentration. This thickness affects GB-mediated plasticity up to 3% strain (Position 1) but has little effect for higher strains where dislocation-dominated plasticity prevails (Positions 2 and 3). At 378 K, GB complexions containing FCC-type and BCC-type SRO are formed. This Ni segregation behavior is consistent with that of different alloys in our previous MC/MD simulation study [35], which found that the GB phase and nano-precipitation tend to disappear at higher temperature due to differing solubility limits that are strongly dependent on the local chemistry of the segregated solute. Although different amounts of SRO can result from varying the solute concentration, the mechanical behavior of SRO complexion alloys is almost insensitive to Ni concentration. Moreover, SRO complexion alloys (annealed at 378 K) sustain higher GB stress during the initial deformation stage and locally produce more heterogeneous GB stress concentrations, relative to the non-SRO complexion alloys (annealed at 913 K).

Amorphous intergranular films have been studied experimentally in Cu-Zr and Al-Ni-

Ce alloys, where they act as sinks for dislocations [15]. This is consistent with the present finding that all stress-strain curves converge to similar behavior in the dislocation-plasticity-dominated regime at high strains (Positions 2 and 3). In Al-Ni-Ce [15] specifically, the stiffness contrast between grain interiors and amorphous GBs was shown to promote homogeneous dislocation emission from the GB network, inhibiting shear localization. The same mechanism appears operative here: Table 1 showed that disordered atoms represent a large GB atom fraction (>84%) in all models, while Table 2 revealed less difference in shear localization between them at Position 2 and larger strains, suggesting that the disordered GB region acted as an effective dislocation sink regardless of Ni concentration and MC/MD temperature, promoting homogeneous dislocation plasticity.

Beyond its role as dislocation sinks, GB thickness also governs the dominant plastic mechanism. Qian et al. [36] demonstrated that increasing amorphous GB thickness shifts the dominant plastic mechanism from dislocation activity to the percolation of shear transformation zones (STZs) within the amorphous phase. The present results are in strong agreement with this picture for two reasons. First, enhanced GB-mediated plasticity without dislocation activity is observed in both Al-2Ni and Al-4Ni with disordered complexions at 913 K, and von Mises stress distributions are more homogeneous in the thicker amorphous complexion of Al-4Ni. Consequently, Al-4Ni is softer than Al-2Ni at Position 1, with more pronounced GB-mediated plasticity. Second, Figure 9 shows direct evidence of STZs in the GB atom region of Al-4Ni (913 K) with thick amorphous complexion.

Differences in shear localization between the two complexion types are most apparent during early deformation stage at 3% strain (Position 1). Al-4Ni at 913 K shows the most severe shear localization at Position 1, accompanied by pronounced early softening. This is consistent with the shear-band-induced softening found in amorphous alloys, which is typically attributed to free volume expansion within shear transformation zones (STZs) [37], a behavior observed in nanocrystalline Ag-Cu alloys with high Cu concentrations [38]. In contrast, alloys with SRO complexions achieve the highest strength regardless of Ni concentration and display significantly less intense shear localization, with clear evidence of stress concentration at GBs (Figure 8). No free-volume expansion was observed in this case, suggesting that this is not the dominant deformation mechanism in nanocrystalline Al-Ni alloys with SRO complexions. Instead, past evidence shows that local heterogeneities induce stress concentrations in amorphous alloys [39,40], consistent with the present results in Figure 8, where high GB stress concentrations are heterogeneously distributed across the GB network.

This shift in behavior, from the STZ-controlled response of amorphous complexion alloys to the stress-concentration-controlled response of SRO complexion alloys, reflects the differing structural character of the two GB types. Heterogeneous GB stresses influence both GB sliding and dislocation emission in nanocrystalline alloys [14]. It is expected that GBs with weaker or absent SRO are more susceptible to homogeneous GB sliding and dislocation emissions from GBs than those with well-developed SRO. In other words, alloys with SRO complexions are governed by the GB stress heterogeneity across the GB network, whereas those with thick amorphous complexions are governed by the fraction and distribution of STZs.

## V. CONCLUSION

Hybrid MC/MD simulations were employed to investigate GB segregation behavior and associated plasticity transitions in nanocrystalline Al-Ni alloys, with emphasis on distinguishing amorphous GB complexions with and without SRO. Two segregation states were identified in nanocrystalline Al-2Ni and Al-4Ni alloys: uniform disordered intergranular film complexions at 913 K and semi-disordered GB complexions containing FCC-type and BCC-type SRO at 378 K.

The two complexion states produced distinctly different mechanical responses. SRO complexions promoted more significant strengthening and reduced shear localization, while disordered intergranular complexions acted as dislocation sinks, effectively enhancing plasticity, but were weaker. This contrast reflects a fundamental shift in governing mechanism: alloys with SRO complexions are controlled by GB stress heterogeneity across the GB network, whereas those with thick amorphous complexions are controlled by the fraction of STZs. These results demonstrate that annealing temperature profoundly influences complexion structure and, consequently, the mechanical behavior of nanocrystalline Al-Ni alloys. Better control over the distribution of SRO phases at GBs within nanocrystalline structures represents a promising avenue for expanding the range of applications of nanocrystalline materials.

**Acknowledgments**

This research received support by the U.S. Department of Energy under grant No. DE-SC0020054 and used resources of the NERSC, a U.S. Department of Energy Office of Science User Facility located at Lawrence Berkeley National Laboratory, operated under

Contract No. DE-AC02-05CH11231.

**Conflict of Interest Statement**

The authors have no conflicts to disclose.

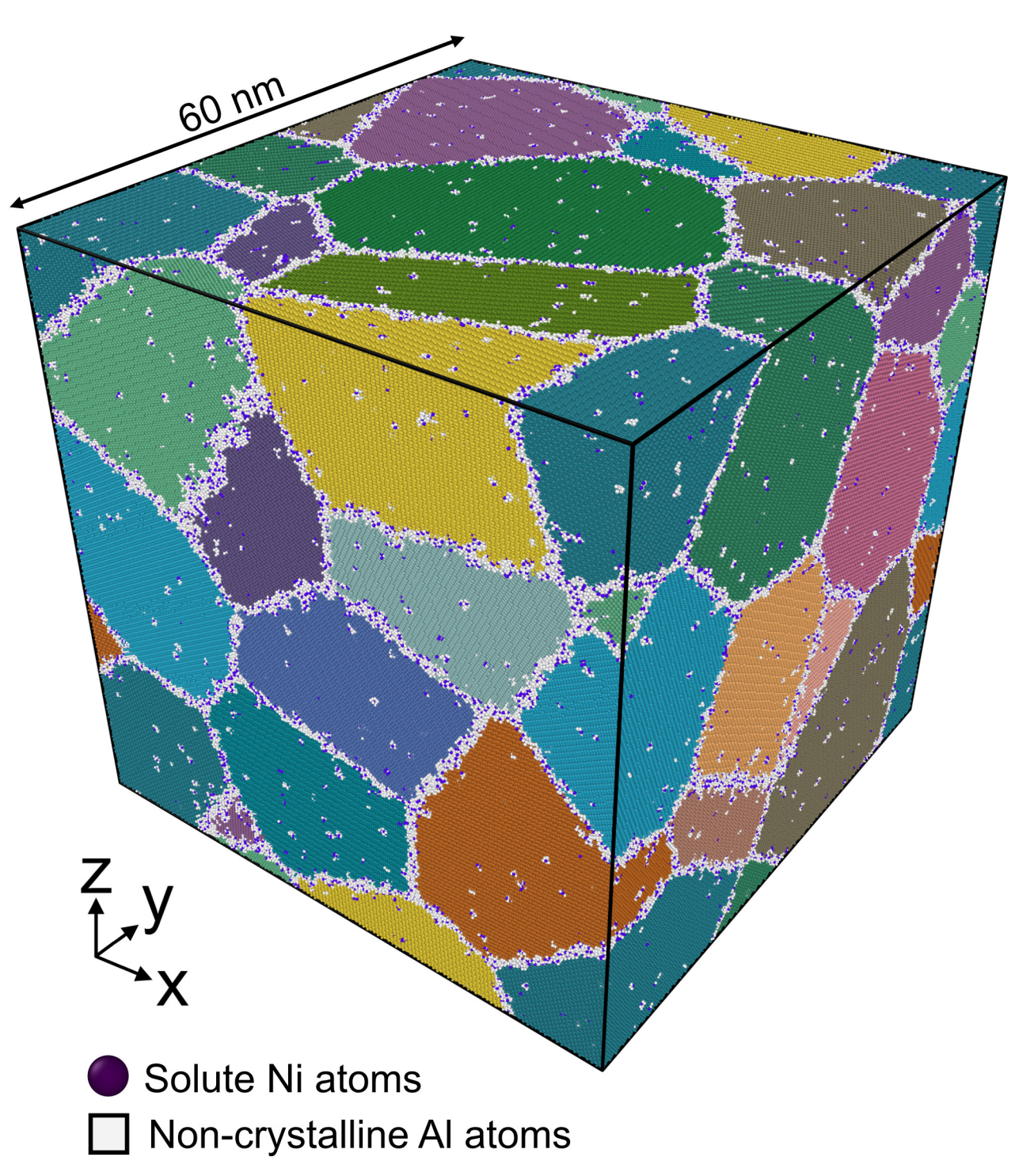

**Figure 1:** Equilibrium microstructure and Ni solute segregation in the Al-4Ni polycrystal predicted by MC/MD simulation at 378 K and then cooled to 300 K. Each grain is highlighted in different colors according to their crystallographic orientation.

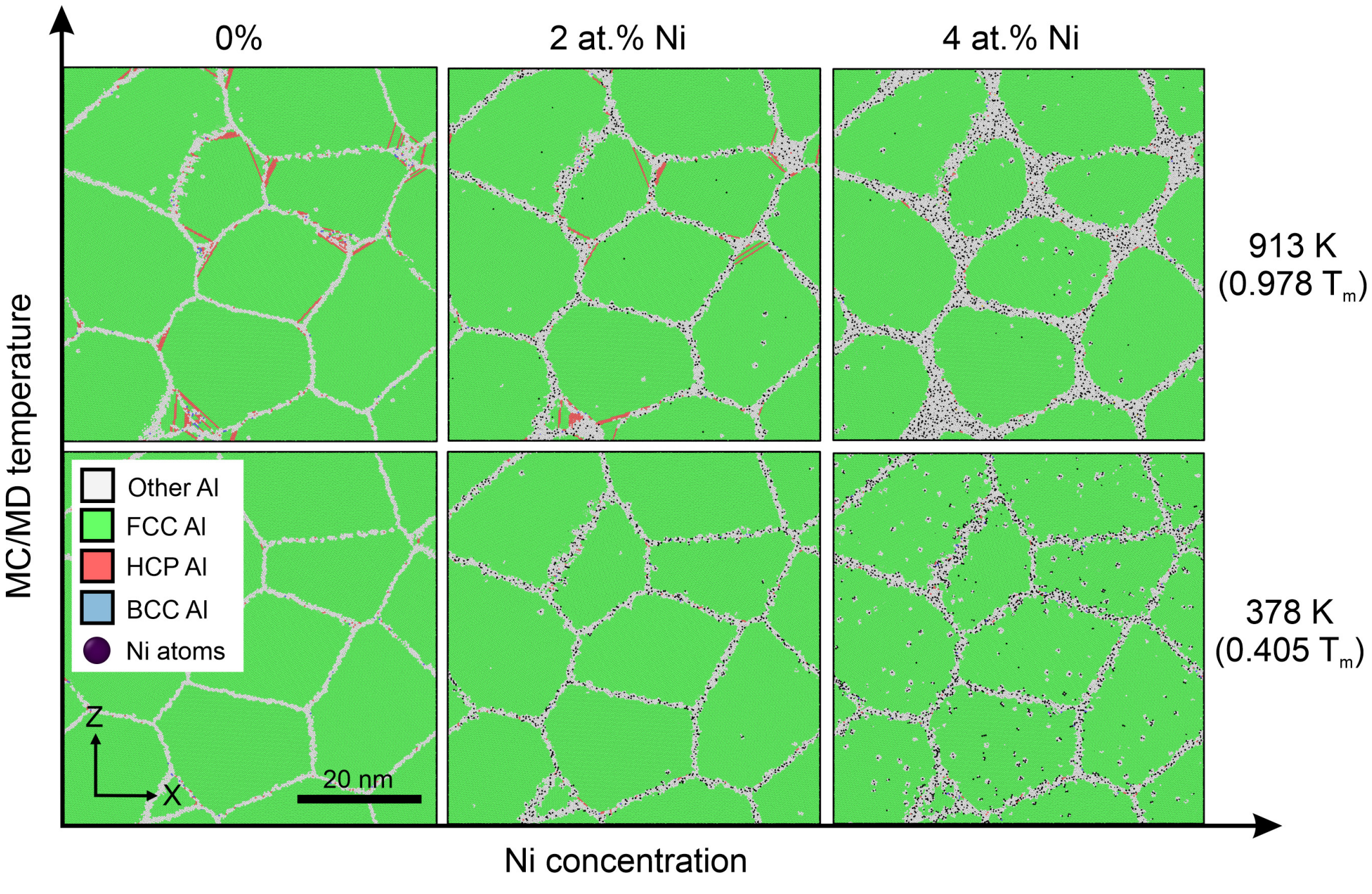

**Figure 2:** Equilibrium Al-Ni phase diagram for total Ni concentrations up to 4 at.% predicted by MC/MD.

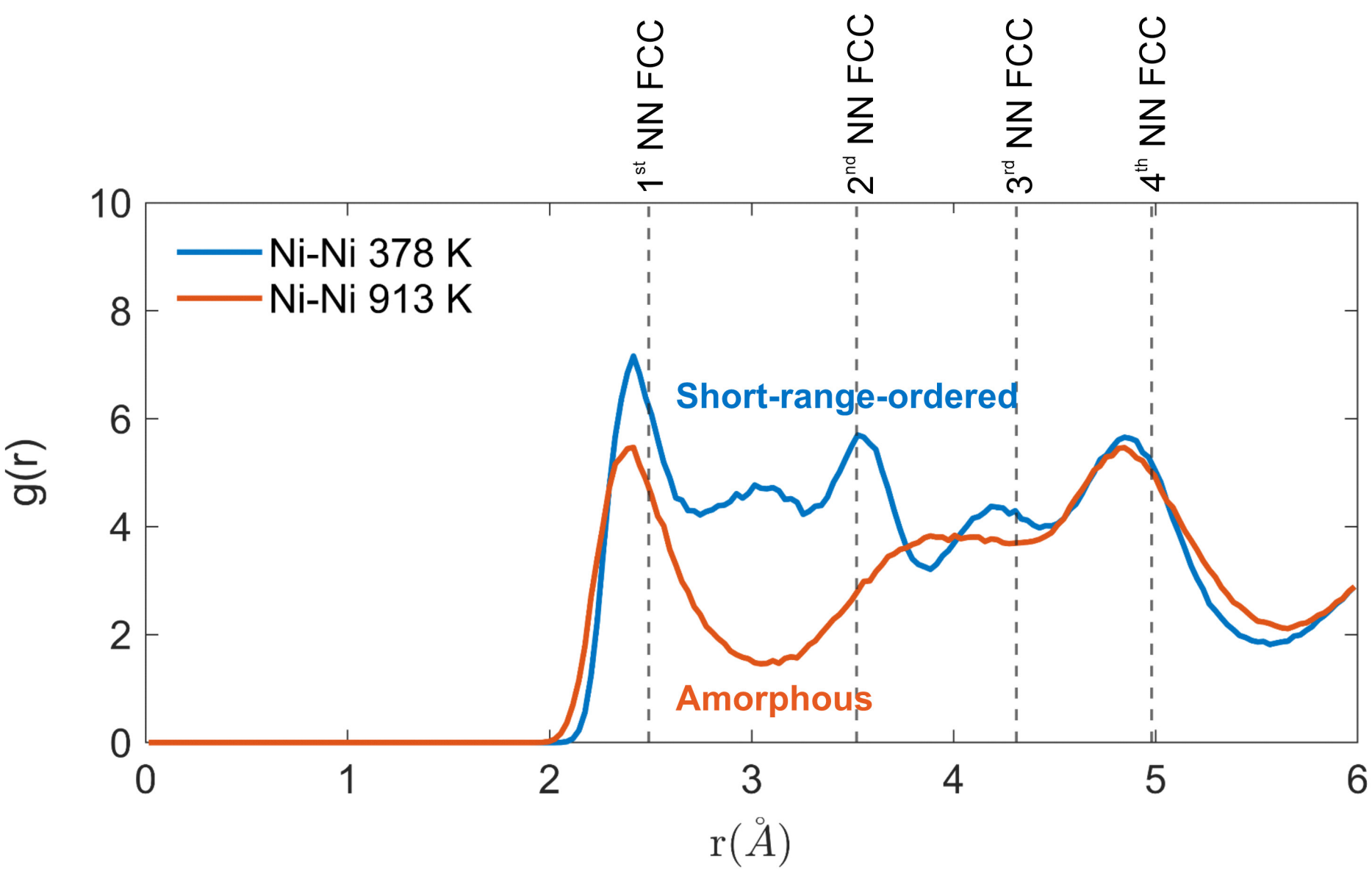

**Figure 3:** Radial distribution function curves for Ni-Ni pairs at grain boundaries in the nanocrystalline Al-4Ni alloy annealed at 913 K and 378 K. The first four nearest-neighbor (NN) distances in a perfect FCC Ni unit cell are indicated by dashed lines.

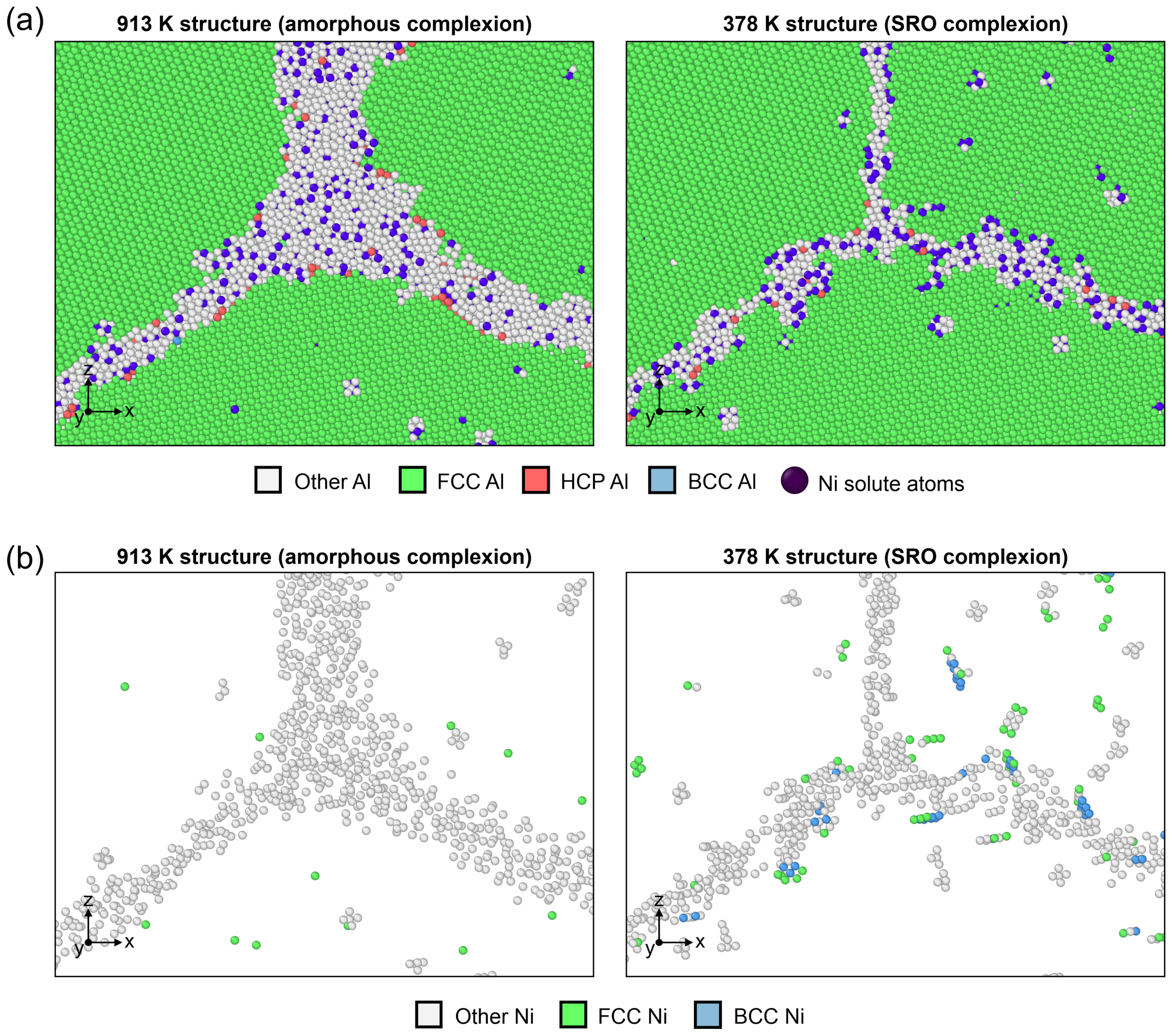

**Figure 4.** Close-up view of the same grain-boundary region in nanocrystalline Al-4Ni annealed at 913 K and 378 K, revealing amorphous and short-range-ordered (SRO) Ni complexions, respectively. Structural analysis shown with (a) all atoms and (b) Ni solute atoms only.

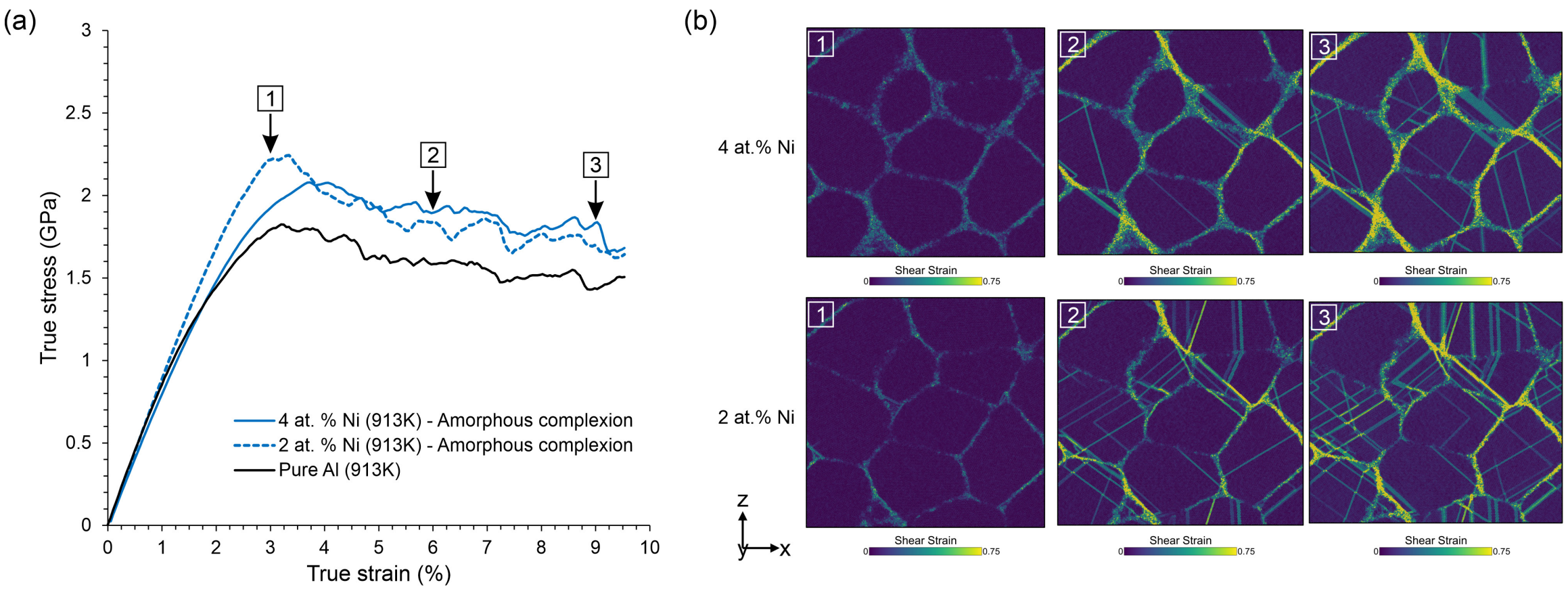

**Figure 5.** Effect of amorphous complexions on deformation in nanocrystalline Al-Ni alloys (annealed at 913 K): (a) simulated tensile stress-strain curves and (b) atomic-scale snapshots of local von Mises shear strain at 300 K.

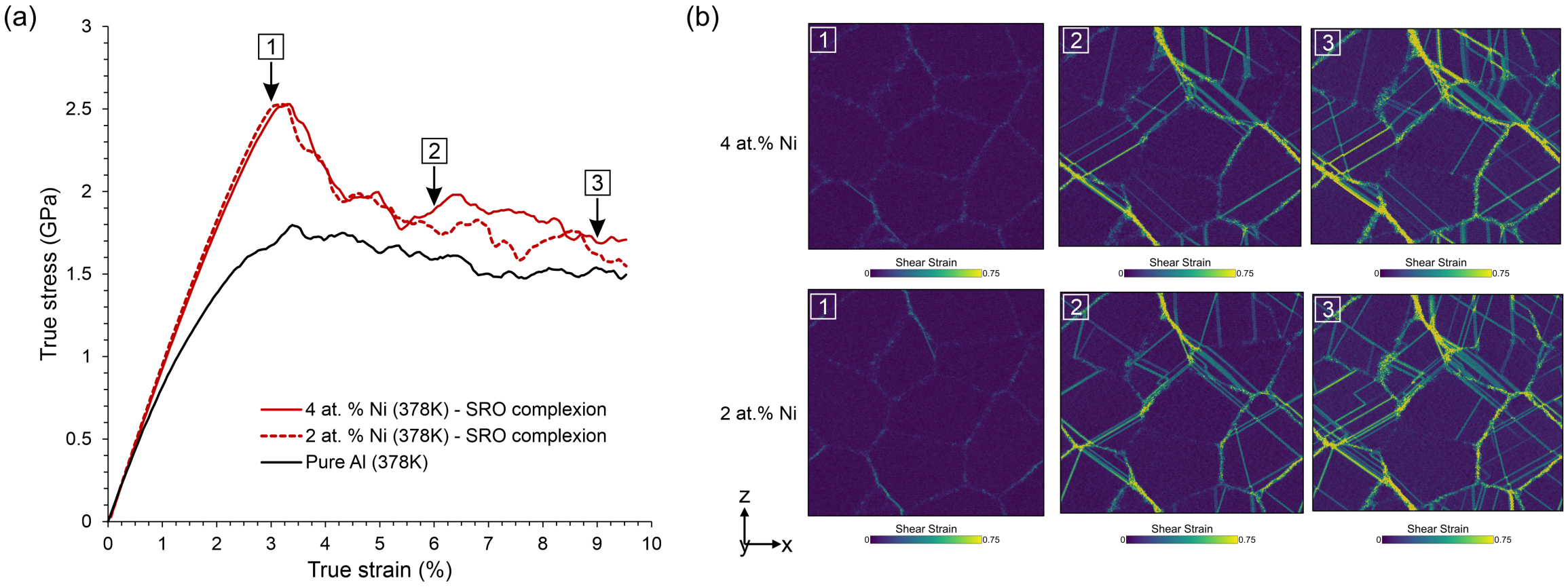

**Figure 6.** Effect of short-range-ordered (SRO) complexions on deformation in nanocrystalline Al-Ni alloys (annealed at 378 K): (a) simulated tensile stress-strain curves and (b) atomic-scale snapshots of local von Mises shear strain at 300 K.

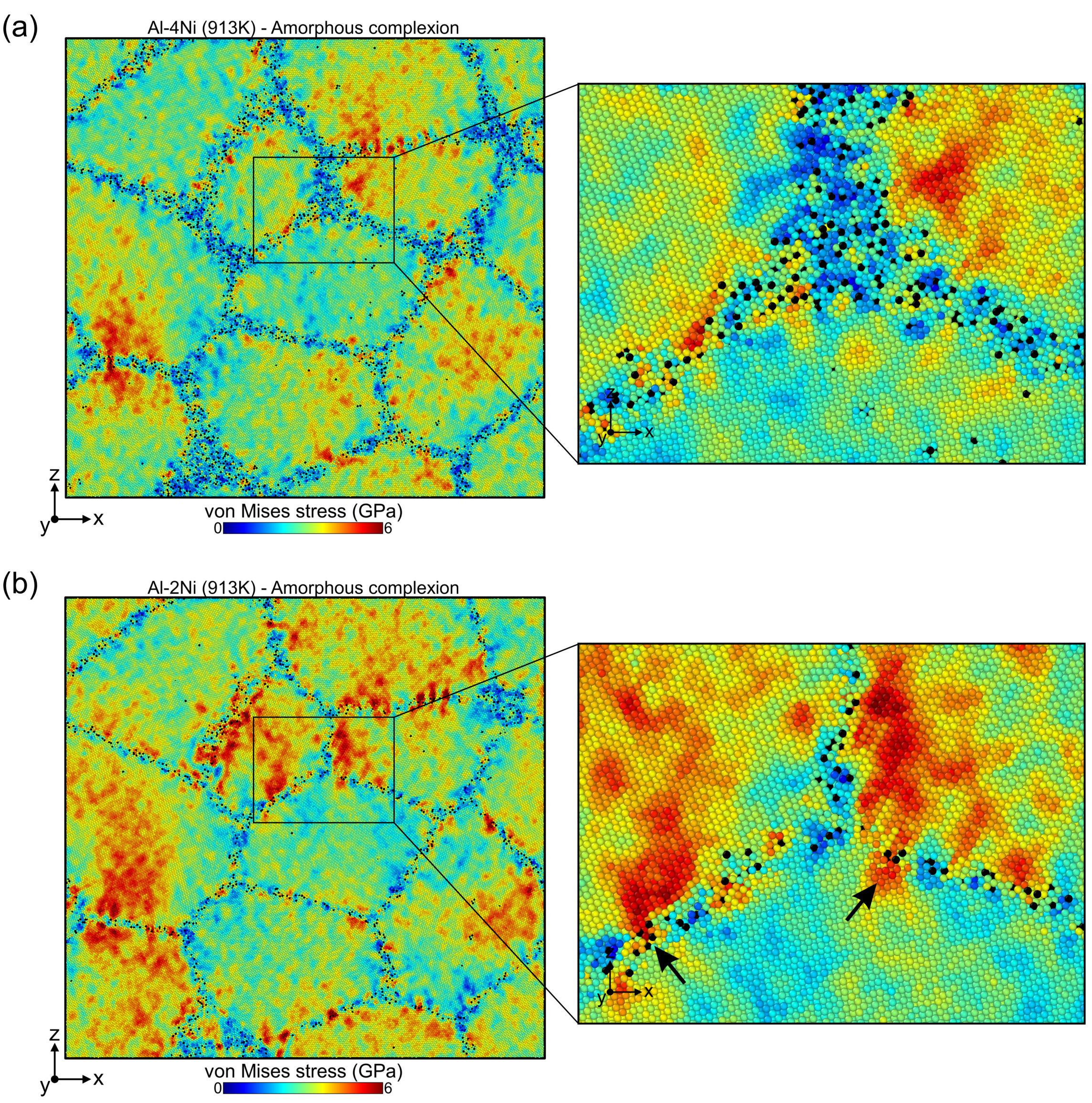

**Figure 7.** Difference in local atomic von-Mises stresses at 3% applied strain at the GB region in Figure 4 with disordered complexions in nanocrystalline (a) Al-4Ni and (b) Al-2Ni alloys annealed at 913 K. Ni solute atoms are highlighted in dark color.

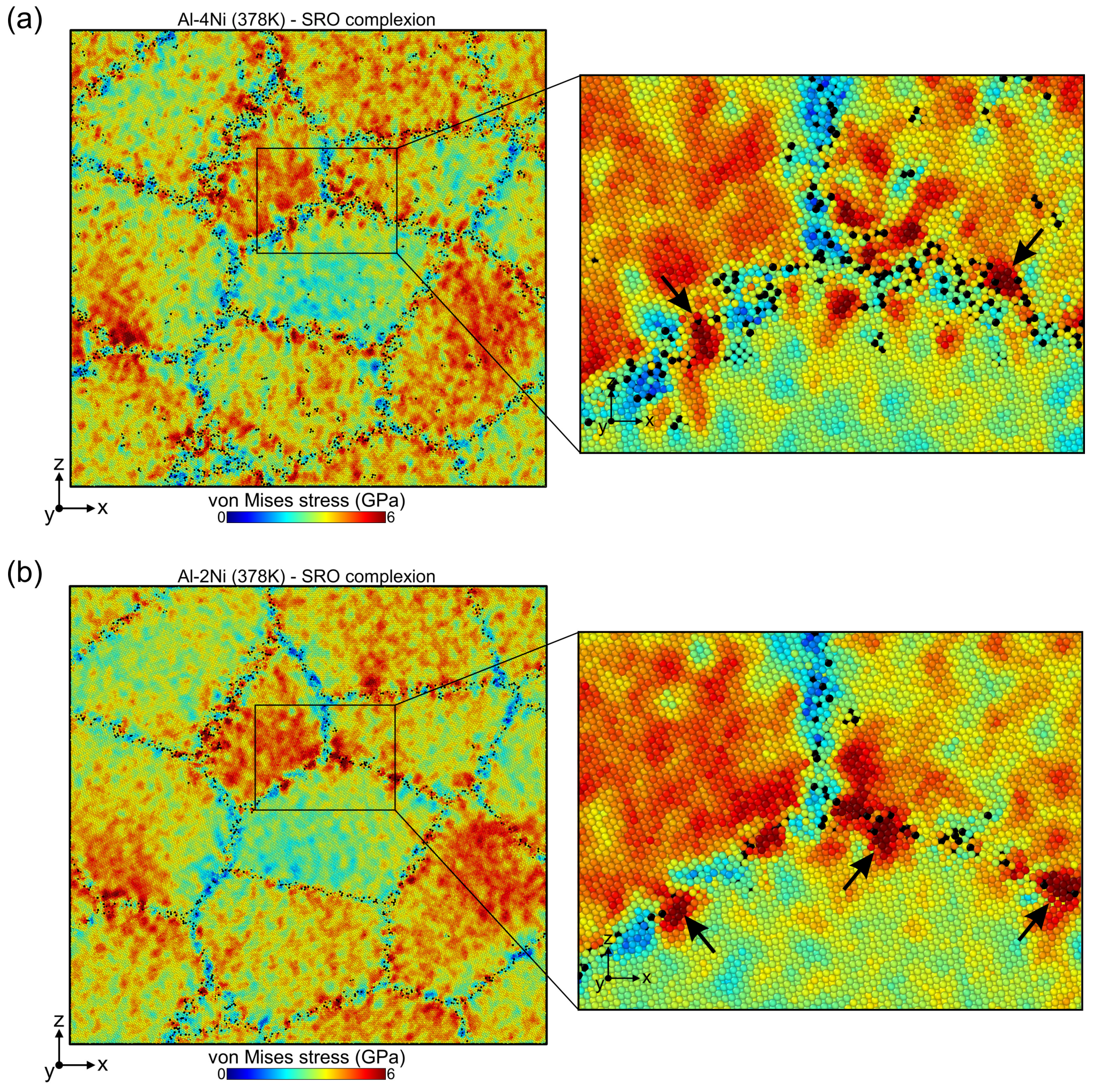

**Figure 8.** Difference in local atomic von-Mises stresses at 3% applied strain (Position 1) at the GB region in Figure 4 with SRO complexions in nanocrystalline (a) Al-4Ni and (b) Al-2Ni alloys annealed at 378 K. Ni solute atoms are highlighted in dark color.

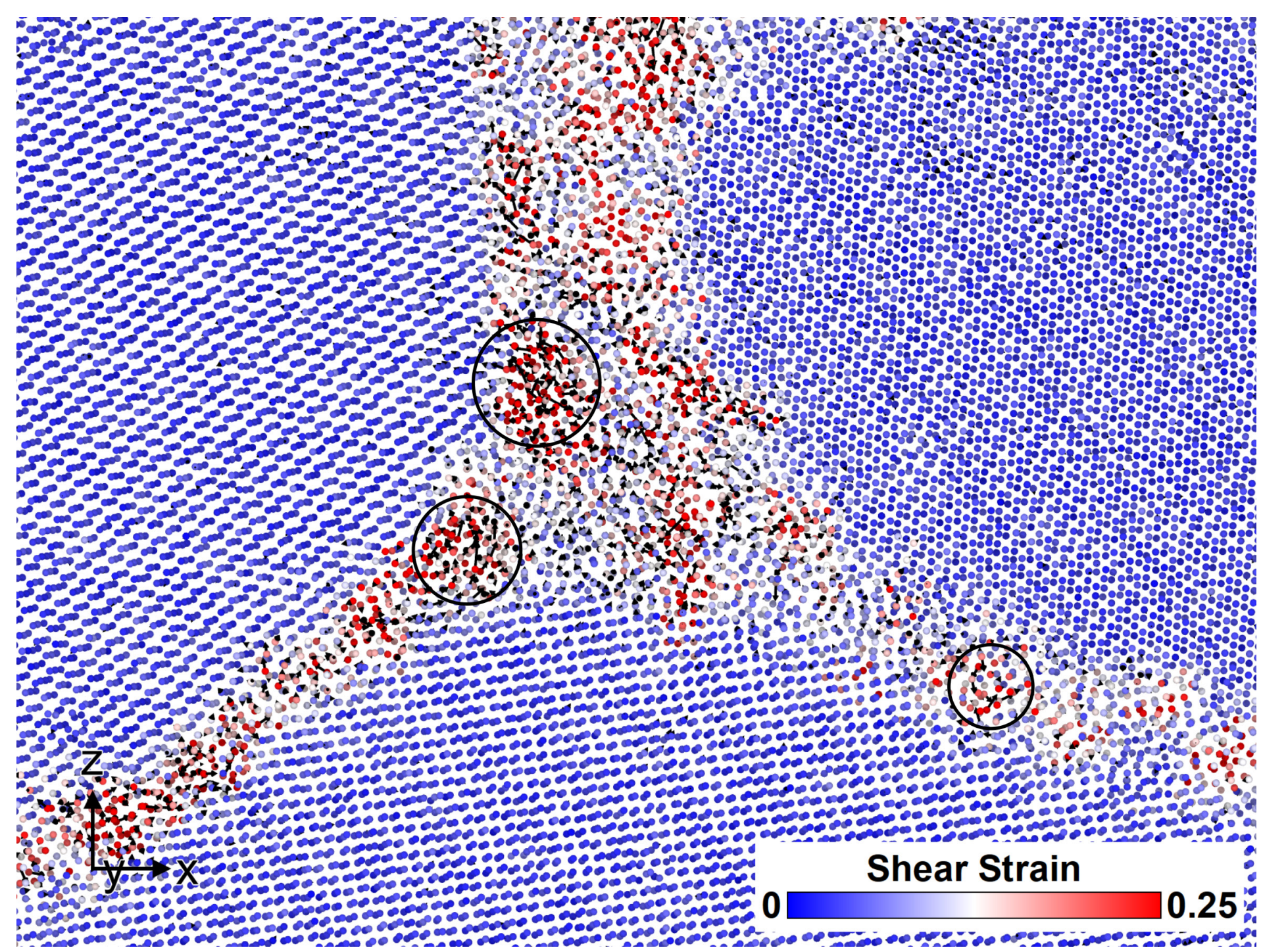

**Figure 9.** Shear transformation zones (STZ, black circles) simulated at 3% strain (Position 1) in the amorphous GB complexion of Al-4Ni (annealed at 913 K) shown in Fig. 7(a). Arrows indicate the atomic displacement direction and magnitude over a 50-ps time increment.

## Highlight Image

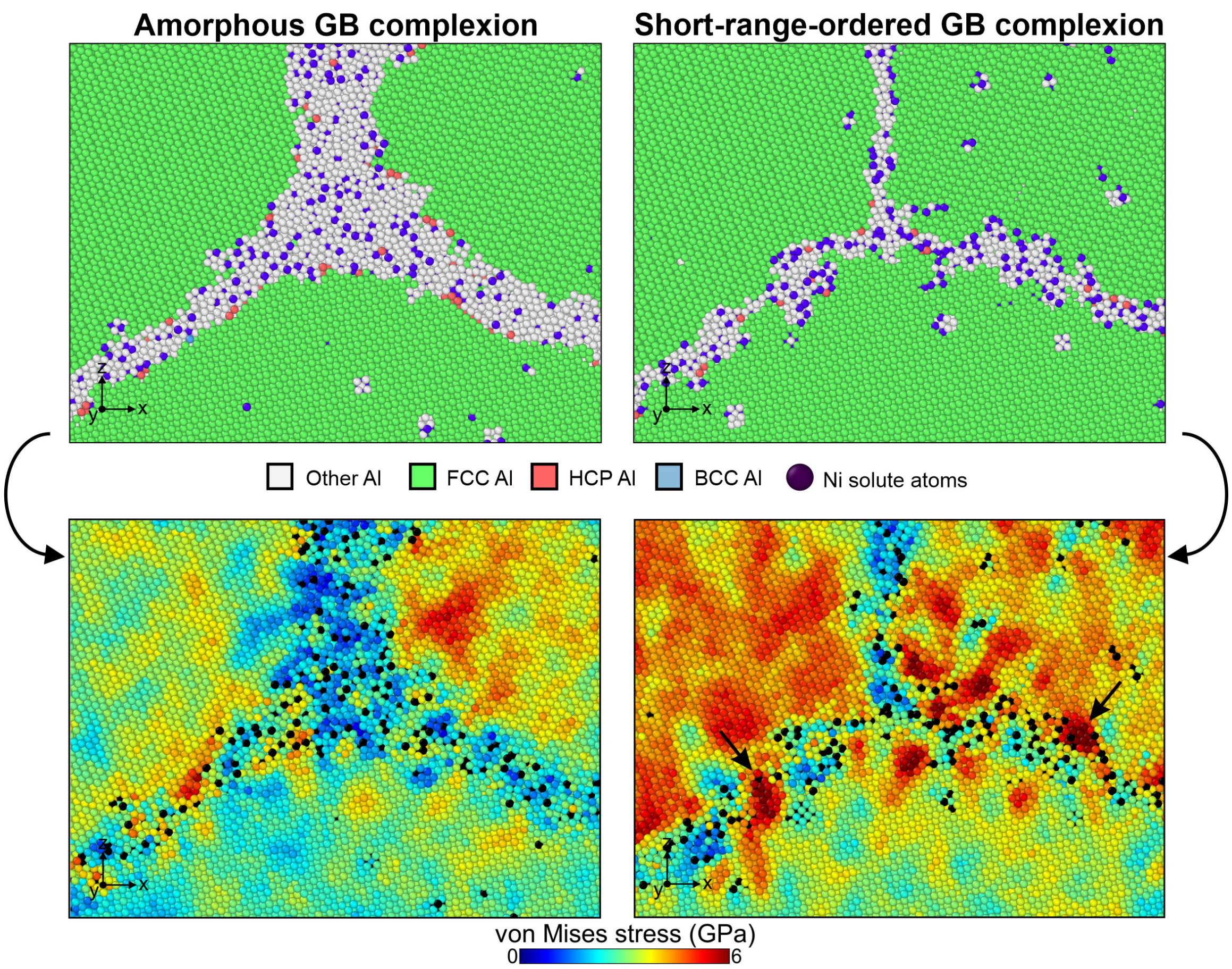